\documentclass[final,authoryear,5p,times,twocolumn]{elsarticle}
\usepackage{fullpage}
\usepackage{subcaption}
\usepackage{pifont}
\usepackage{latexsym}
\usepackage{mathrsfs}
\usepackage{amssymb,amsbsy}
\usepackage{amsfonts}
\usepackage{graphicx}
\usepackage[colorinlistoftodos]{todonotes}
\usepackage{varwidth}
\usepackage{algorithm}
\usepackage{algorithmic}
\usepackage{lmodern}
\usepackage{natbib}
\usepackage{mathtools}      
\usepackage[titletoc]{appendix}
\usepackage{titletoc}
\usepackage{ntheorem}

\newcommand{\astrosat}{{\it AstroSat }}
\newcommand{\jude}{{\it JUDE }}

\journal{Astronomy and Computing}

\begin{document}

\begin{frontmatter}

\title{JUDE: An Ultraviolet Imaging Telescope Pipeline}
\author[1]{Jayant Murthy}
\author[2]{P. T. Rahna}
\author[1]{Firoza Sutaria}
\author[3]{Margarita Safonova\corref{cor1}}
\ead{margarita.safonova62@iiap.res.in}
\author[2]{S. B. Gudennavar}
\author[2]{S.~G.~Bubbly}

\cortext[cor1]{Corresponding author}
\address[1]{Indian Institute of Astrophysics, Bengaluru}
\address[2]{Department of Physics, Christ University, Bengaluru}
\address[3]{M.~P.~Birla Institute of Fundamental Research, Bengaluru}

\begin{abstract}
The Ultraviolet Imaging Telescope (UVIT) was launched as part of the multi-wavelength Indian \astrosat mission on 28 September, 2015 into a low Earth orbit. A 6-month performance verification (PV) phase ended in March 2016, and the instrument is now in the general observing phase. UVIT operates in three channels: visible, near-ultraviolet (NUV) and far-ultraviolet (FUV), each with a choice of broad and narrow band filters, and has NUV and FUV gratings for low-resolution spectroscopy. We have written a software package (\jude) to convert the Level~1 data from UVIT into scientifically useful photon lists and images. The routines are written in the GNU Data Language (GDL) and are compatible with the IDL software package. We use these programs in our own scientific work, and will continue to update the programs as we gain better understanding of the UVIT instrument and its performance. We have released \jude under an Apache License.
\end{abstract}


\end{frontmatter}

\section{Introduction}

The Ultraviolet Imaging Telescope (UVIT) was first proposed as part of the multi-wavelength \astrosat mission in the late $20^{th}$ century \citep{Pati1998,Pati1999} and was launched on 28 September, 2015 into a low Earth orbit (650 km, $6^{\circ}$ inclination) by an ISRO (Indian Space Research Organization) PSLV (Polar Satellite Launch Vehicle) launcher. The UVIT payload doors were opened on 30 November, 2015 and a 6 month performance verification (PV) phase began, ending in March 2016. The ground calibration of the payload has been discussed by \citet{Postma2011} with in-flight tests by \citet{Subramaniam2016}. 

There are now a complete set of PV phase observations which have been used to characterize the instrument \citep{Tandon2016} with in-flight calibration and verification done by \citet{Tandon2017} and \citet{Rahna2017}, and data are being released to the observers. However, there are still (at the date of writing) issues with the UVIT pipeline and there are data sets which cannot be processed. More importantly from our viewpoint is that the source code is proprietary and difficult to modify.

We have chosen to create an alternative set of routines ({\it JUDE}: Jayant's UVIT Data Explorer) released under an Apache License\footnote{\textit{http://www.apache.org/licenses/LICENSE-2.0}} which may be freely used and modified. \jude  begins with the Level 1 data provided by the Indian Space Science Data Centre (ISSDC) and produces photon lists and images suitable for scientific analysis.

It should be stressed that this package is not intended to be a replacement for the official UVIT pipeline but is rather a tool to examine the data in more detail. However, the authors are using {\em JUDE} for their own scientific purposes, and are actively maintaining and modifying the software as the instrument characterization improves. \jude is archived at the Astrophysics Source Code Library \citep{Murthy_ascl} and the latest version is available on GitHub (https://github.com/jaymurthy/JUDE). We solicit feedback on its operation and further improvements.

\section{Instrumentation and Observations}

The UVIT instrumentation has been described in \citet{Tandon2016}, with a description of the entire \astrosat mission by \citet{Singh2014}. The UVIT instrument consists of two 35-cm Ritchey-Chr\'{e}tien telescopes with three intensified CMOS detectors. One telescope feeds a far ultraviolet (FUV) detector (1300 -- 1800~\AA),  with the other feeding two detectors in the near ultraviolet (NUV: 1800 -- 3000~\AA) and visible (VIS: 3200 -- 5500~\AA), respectively, through a dichroic filter. A filter wheel in front of each detector allows different spectral ranges to be selected. The visible channel is not intended for science purposes and is used solely to track the spacecraft motion. Further information about the \astrosat mission and the instruments aboard may be obtained from the \astrosat Science Support Cell (http://astrosat-ssc.iucaa.in).

The UVIT instrument has observed a number of targets during the PV phase, where an observation may be broken into several exposures of several hundred seconds in length. UVIT has now begun regular observations and is shifting to an observer driven mission with proposals from the Indian and international astronomical community. The official UVIT pipeline is tasked with producing scientifically usable images, including flat-fielding, distortion correction, drift correction, and absolute calibration from the raw data, and is available from the Science Support Cell referenced above.

\section{\jude}
\subsection{Overview}

We have written the {\em JUDE} software system entirely in the GNU Data Language \citep{GDL2010}. GDL was chosen for the development environment because it is an interpreted language which lends itself to the interactive analysis of data. This was invaluable in the development of the pipeline where we could check each step and run commands interactively. GDL is an open source version of the Interactive Data Language\footnote{http://www.harrisgeospatial.com} (IDL) and will run on a wide variety of systems, as will IDL. It will run most IDL programs without modification (and {\it vice versa}) allowing access to the rich library of utilities developed for IDL over the last four decades, easing the development of {\it JUDE}. We have tested \jude using both GDL (version 0.9.6) and IDL with identical results on multiple operating systems. In the remainder of the paper, GDL and IDL may be freely interchanged.

\begin{table*}[t]
\centering
\caption{JUDE modules}
\label{tab:modules}
\begin{tabular}{|l|l|}
\hline
Module & Purpose\\
\hline
\multicolumn{2}{|c|}{Accessory Programs}\\
\hline
jude\_get\_files & Returns names of data files.\\
jude\_params & Sets operating parameters.\\
jude\_err\_process & Error handler.\\
\hline
\multicolumn{2}{|c|}{Level 1 Data}\\
\hline
jude\_read\_vis & Reads the visible data.\\
jude\_create\_uvit\_hdr & Creates FITS data header for Level 2 data files.\\
jude\_read\_hk\_files & Reads housekeeping files.\\
jude\_set\_dqi & Checks instrumental parameters.\\
jude\_get\_xy & Extracts individual events from Level 1 data.\\
jude\_cnvt\_att\_xy & Calculates $X$ and $Y$ shifts from boresight.\\
jude\_check\_bod & Check and reject bright object detection (BOD).\\
\hline
\multicolumn{2}{|c|}{Level 2 Data}\\
\hline
jude\_register\_data & Corrects for spacecraft motion.\\
jude\_add\_frames & Combines individual frames into image.\\
jude\_vis\_shifts & Calculates spacecraft motion from visible data.\\
jude\_add\_vis & Adds visible frames together.\\
jude\_obs\_log & Creates observation log from Level 2 files.\\
jude\_merge\_files & Combines files with overlapping data.\\
jude\_match\_vis\_offsets & Matches the visible offsets to the UV channels.\\
jude\_apply\_cal & Applies photometric calibration to data\\
jude\_centroid & Tracks spacecraft motion through centroids of stars.\\
\hline
\multicolumn{2}{|c|}{Driver Routines}\\
\hline
jude\_driver\_vis & Chains individual programs to produce Level 2 files in the VIS.\\
jude\_driver\_uv  & Chains individual programs to produce Level 2 files in the UV.\\
jude\_uv\_cleanup & Post processing of Level 2 data.\\
jude\_interactive & Interactive exploration of Level 2 data.\\
\hline
\hline
\end{tabular}
\end{table*}

The two UV channels typically operate in photon counting mode with the VIS camera operating in integrating mode. The primary science from UVIT will come from the two UV imagers with the VIS camera intended solely to correct for spacecraft motion. We have therefore focussed on producing scientific products from the two UV channels.

Our starting point is the Level 1 data provided by the ISSDC from which we produce photon lists and images of the sky which may be used for further scientific analysis (Level 2 data). We have tested the \jude system by reducing the entire set of PV observations and are now using it to analyse our GT (Guaranteed Time) observations. In the following sections, we will first describe the Level 1 and 2 data files, the operation of the pipeline and the logic behind the primary components of {\it JUDE}. The individual modules in \jude are listed in Table~\ref{tab:modules}. Each module is self-documenting with a prologue containing a description of the inputs and outputs. We use a number of routines from the IDL Astronomy Library \citep{Landsman1995} without modification  (\ref{app:external_modules}: Table~\ref{tab:idlastron}) and the MPFIT routines \citep{mpfit2009} for PSF fitting.

\subsection{Level 1 Input Files}

Data from the \astrosat spacecraft are transmitted to the \astrosat Data Centre at the ISSDC where they are separated by instrument and written into Level 1 data files. They are then sent to the Payload Operations Centre (POC) at the Indian Institute of Astrophysics (IIA) where they will be processed by the official UVIT pipeline into Level 2 data products, suitable for scientific analysis. After validation, the Level 2 products will be sent back to the ISSDC which has the primary responsibility for data archival and dissemination.

The Level 1 data are distributed as a single zipped archive (earlier tar.gz) for each observation. The files within the archive are in a number of subdirectories organized by orbit number and type of file, all under a single top level directory. All data files are FITS binary tables \citep{FITS2010} with individual frames tagged by mission time and containing, depending on the file, either housekeeping or science data. Due to the manner in which the data are read and transmitted to the ground, an archive may consist of multiple housekeeping and science files, possibly with duplicated frames and/or broken across multiple files. 

We extracted all the files from the archive using unzip (or tar -xf) into a single top-level directory, which we use four sets of files (irrespective of subdirectory): 
\begin{enumerate}
\item Level 1 data files for VIS (``uvtV''): Used to correct for spacecraft motion. Not required for pipeline operation.
\item Level 1 data files for either FUV (``uvtF'') or NUV (``uvtN''): At lease one file must be present.
\item Housekeeping files (``*.lbt''): Frames with no housekeeping information are rejected. 
\item Attitude files of spacecraft boresight(``*.att''):  Required for a preliminary estimate of pointing.
\end{enumerate}

\subsection{Running \jude}

\subsubsection{Pipeline Operation}

We have written \jude as a set of procedures which must be run from within the GDL (or IDL) language. Each procedure is designed to be run independently but will operate as a pipeline if chained together. We have written a shell program (\ref{app:running_the_pipeline}) which we have used to process the entire set of PV observations without active monitoring. These files may be further processed by the user to co-add different exposures of a single target, extract point sources, and for other scientific purposes. 

The individual \jude procedures are listed in Table \ref{tab:modules} and their operation is controlled by a GDL data structure (\ref{app:params}) where the structure is defined in, and defaults are set by, {\it jude\_params.pro}. The user may change the parameters on a global level by editing the parameter file or on the fly through the command line.

\subsection{VIS Processing}\label{sec:vis_proc}

\begin{figure}[t]
\includegraphics[width=3in]{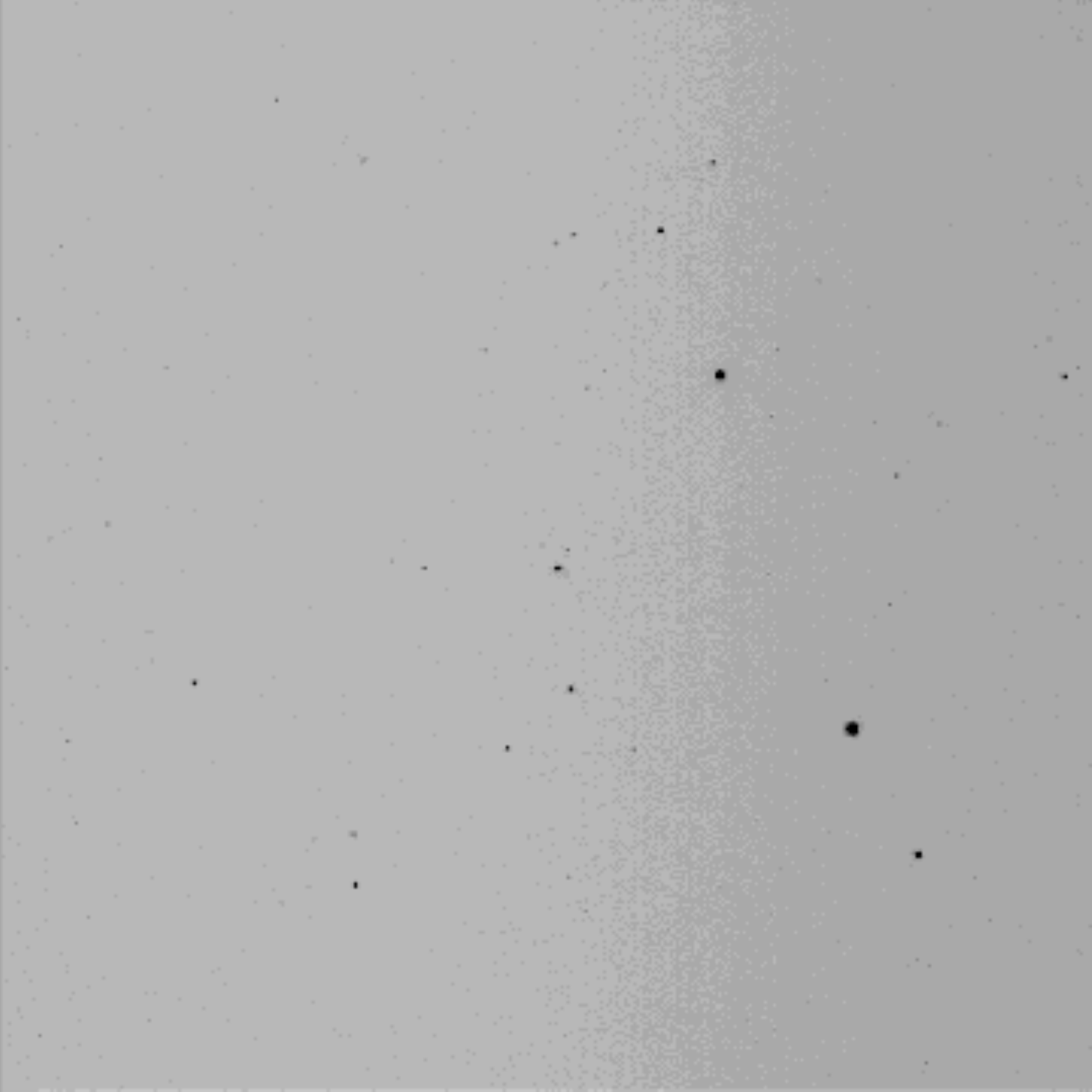}
\caption{A 1 second frame of Holmberg II with the VIS imager.}
\label{fig:vis_image}
\end{figure} 

Observations in the VIS channel are taken in integration mode where the $512\times 512$-pixel CMOS chip is read once per second and the pixels stored in a binary FITS table where the data from each frame are in 261 rows of 1008 pixels each, first with increasing $x$ (0 -- 511) and then increasing $y$ (0 -- 511). The  final 944 ($ 261 \times 1008 - 512 \times 512$) elements are left blank and should be skipped. We used {\it jude\_read\_vis.pro} to read each VIS frame (Fig.~\ref{fig:vis_image}) and extracted the $x$ and $y$ positions of the stars in the field with the library routine {\it find.pro} (based on DAOPHOT: \citet{Stetson1987}). The spacecraft motion is just the shifts between successive frames which we calculated and wrote into a text file using {\it jude\_vis\_shifts.pro}. 

\begin{figure*}[t]
\includegraphics[width=3in]{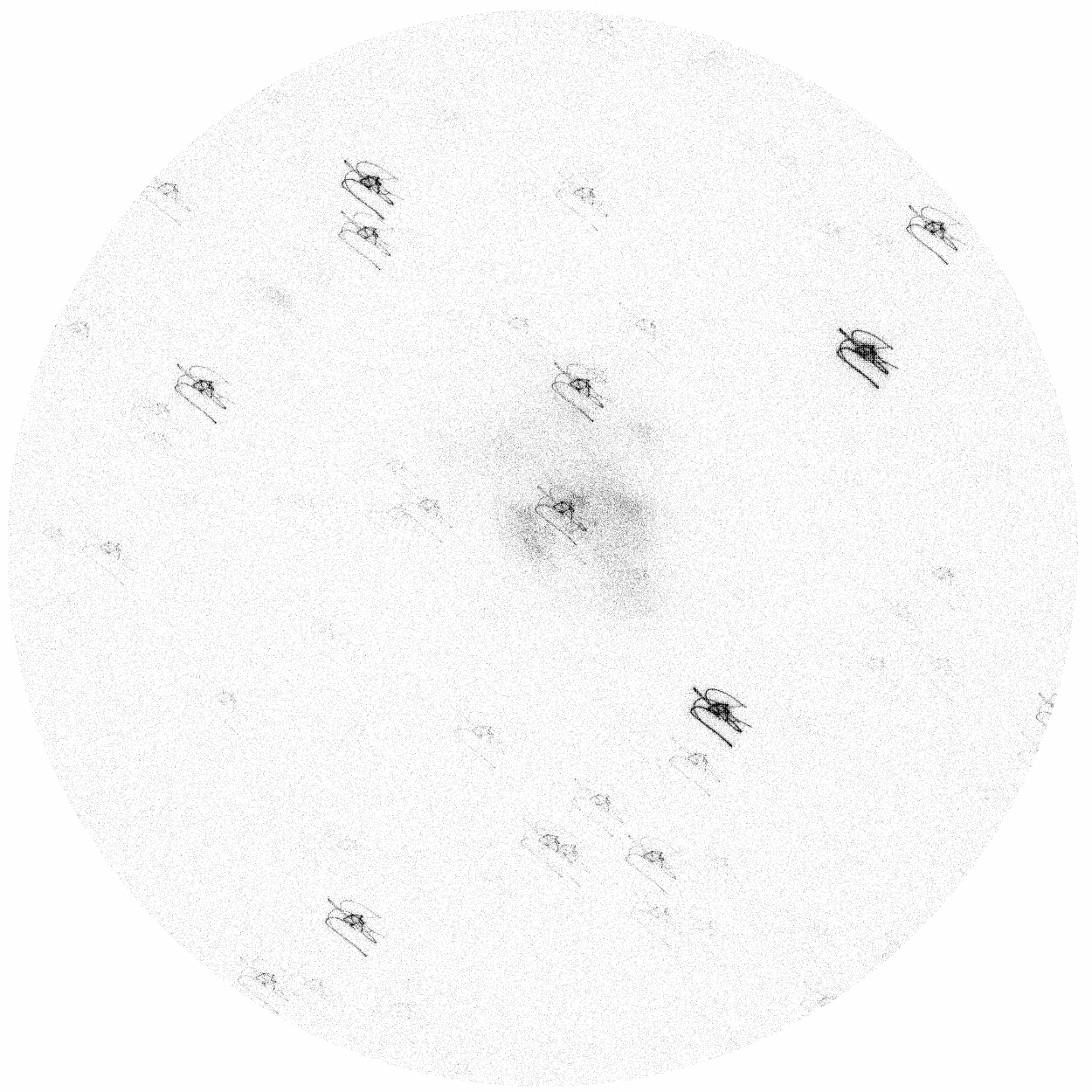}
\includegraphics[width=3in]{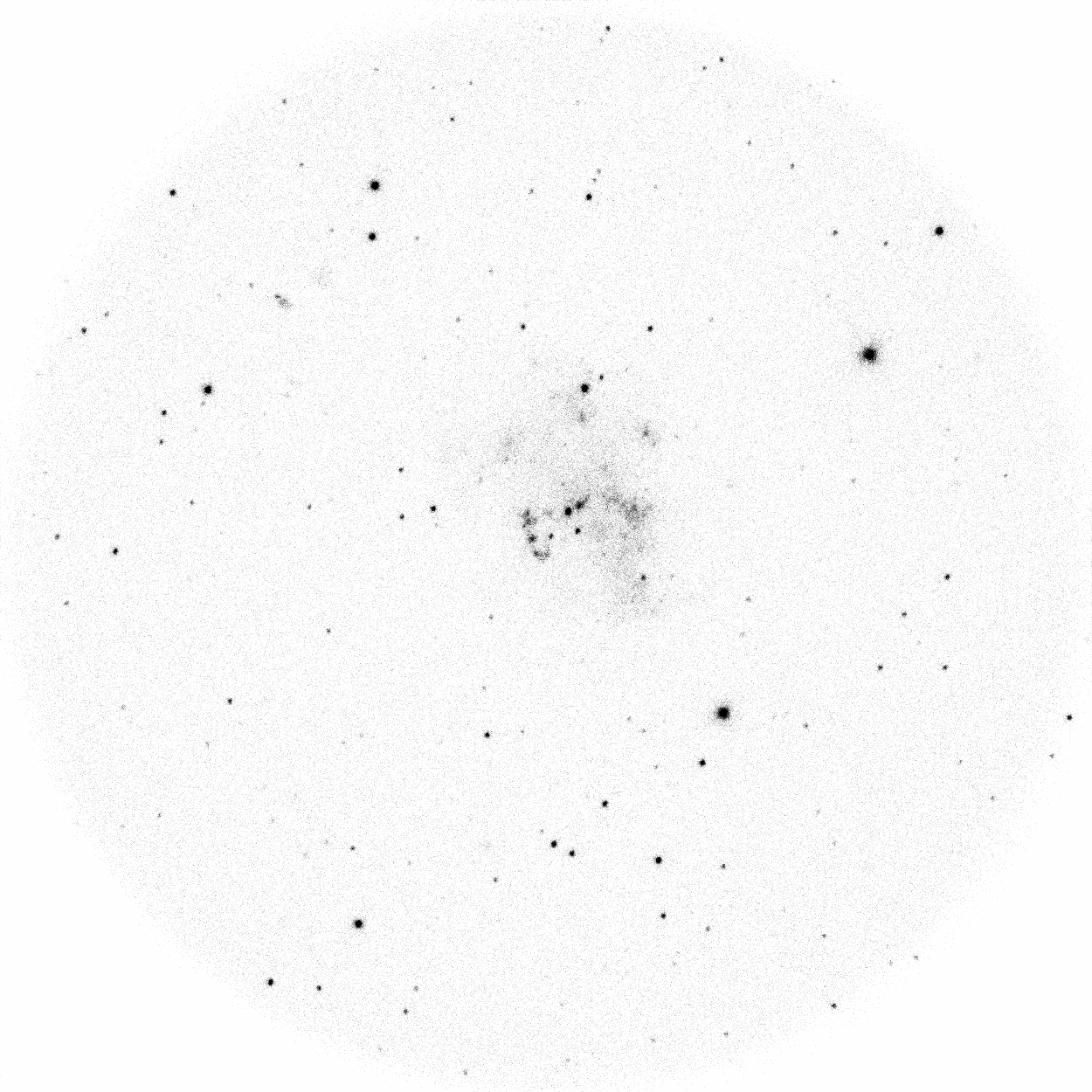}
\caption{Holmberg II without correction for spacecraft motion on the left and after correction with the VIS offsets on the right.}
\label{fig:vis_correction}
\end{figure*}

The NUV and VIS channels share a telescope with the NUV light reflected, and the VIS light transmitted, through a dichroic and the shifts for the NUV image are mirrored around the $x$-axis 
\begin{align}
&x_{\rm\small NUV} = x_{\small\rm VIS}\,,\nonumber \\
&y_{\small\rm NUV} = -y_{\small\rm VIS}\,.
\label{nuv_rot}
\end{align}
The FUV channel has its own telescope, and we found (empirically) that it is rotated from the VIS by $35^{\circ}$, 
\begin{align}
&x_{\small\rm FUV} = x_{\small\rm VIS} \cos{(35^{\circ})} - y_{\rm VIS} \sin{(35^{\circ})}\,,\nonumber \\
&y_{\small\rm FUV} = x_{\small\rm VIS} \sin{(35^{\circ})} + y_{\small\rm VIS} \cos{(35^{\circ})}\,.
\end{align}
We use {\it jude\_match\_vis\_offsets.pro} to store the offsets into the UV channels, using the spacecraft time as a reference. We have shown the original image with the spacecraft motion and the corrected image in Fig.~\ref{fig:vis_correction}. In practice, we find, as we will show below, that we can obtain better resolution through self-registration, {\it i.e.} using stars in the NUV (FUV) channel to correct for the motion in the NUV (FUV) channel.   

\subsection{Housekeeping and Attitude}

\begin{table}
\caption{DQI values}
\label{tab:dqi}
\begin{tabular}{lll}
\hline
DQI & Description & Module  \\ 
\hline
0	&	Data valid. & \\
1	& BOD & {\it jude\_check\_bod} \\
2	& Unused. 						&   						\\
4   & Unused.               		&   						\\
8   & No HK data. 		& {\it jude\_set\_dqi} \\
16  & Duplicate frames.	    		& {\it jude\_get\_xy}  \\
32  & Filter closed/invalid.  	& {\it jude\_set\_dqi} \\
64  & Voltage out of bounds.   	& {\it jude\_set\_dqi} \\
128 & $>1000$ counts. 		& {\it jude\_get\_xy}	\\
1024& Parity violation. & {\it jude\_get\_xy} \\ \hline
\end{tabular}
\end{table}
The housekeeping files (*.lbt) contain a wealth of information about the health and the environmental conditions of the UVIT instrument, most of which are not relevant to our scientific analysis. The attitude files (*.att) contain the spacecraft boresight position as determined by the star sensor approximately every 16 seconds. We use {\it jude\_read\_hk.pro} to read the housekeeping and attitude files and {\it jude\_set\_dqi.pro} to match the housekeeping and attitude information with the sensor data based on the mission time. We perform basic sanity checks as part of our Level 1 processing and set the DQI (data quality index) as in Table~\ref{tab:dqi}. Note that the DQI values are powers of two and additive; it is trivial to diagnose which data were rejected and why.

A bright object detection (BOD) sequence was run at the beginning of every observing sequence in which the gain was ramped up to the operating value. A few frames were sampled at each gain to ensure that there was no potential for damage to the instrument. Not all data files contain the BOD due to the manner in which the data were transmitted to the ground but we checked every observation and flagged those that were part of the BOD sequence.

\subsection{Processing UV Data}

The UVIT detectors are intensified $512\times 512$ CMOS detectors \citep{Tandon2016} read 29 times a second with an exposure time of 0.035 seconds per frame. Individual photons fall on the photocathode where they eject electrons that are accelerated through a microchannel plate (MCP) onto a phosphor screen. The resulting flashes of green light, each of which represents a single photon hit, are recorded on the CMOS sensor.  On-board software converts these flashes into individual photon events with a nominal resolution of $\frac{1}{8}$ of a pixel and packs each event into a byte array which is stored on-board and ultimately transmitted to the ground. The ISSDC separates the data by instrument and channel and passes them on to the user as Level 1 data files.

There is another mode in which a small ($100\times 100$ pixel) window is read at 600 frames per second as well as a transmission grating which produces low resolution spectra. The processing scheme is the same for those modes and we do produce usable data but have not tested the data products for scientific integrity.

\begin{figure}[t]
\includegraphics[scale=0.35]{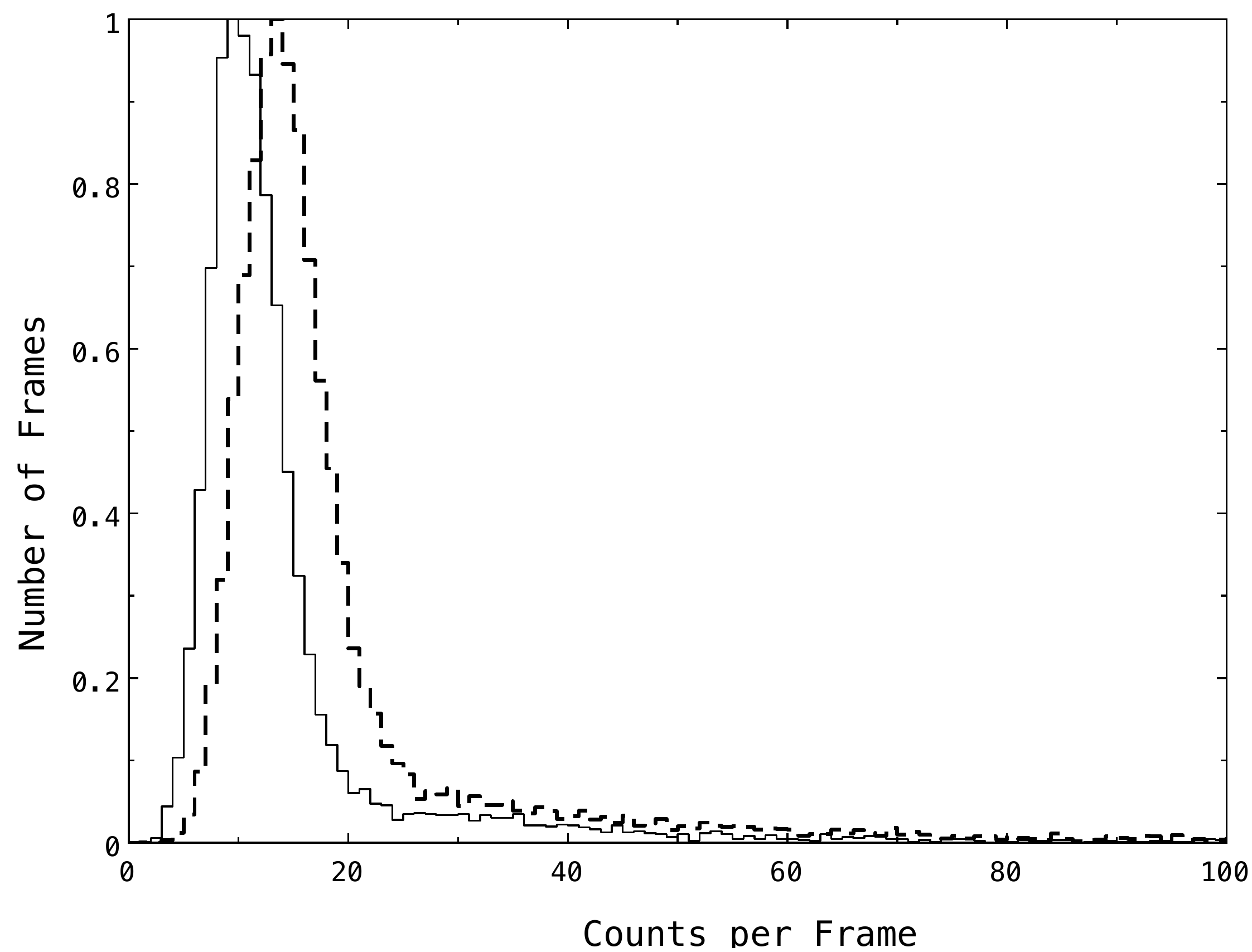}
\caption{Normalised histogram of the count rate per frame in the FUV (solid line) and NUV (dashed line).}
\label{fig:event_histogram}
\end{figure}

Each event in the Level 1 data is stored as 6 bytes with two bytes for each of the $x$ and $y$ positions and the third for diagnostic information about that event (not used in {\it JUDE}). One frame contains a maximum of 336 events with further events resulting in a new frame. We read each frame and extracted the photon events using {\em jude\_get\_xy.pro}. The CMOS sensors are subject to cosmic ray hits giving rise to Cherenkov radiation within the MCP. This blue light results in a huge splash on the detector which is interpreted as a large number of individual photon hits by the on-board centroiding algorithm, all spurious. The actual count rate in one of our observations is plotted in Fig.~\ref{fig:event_histogram} where the cosmic ray hits are represented by the long tail at higher count rates. The effect of including the spurious counts is to increase the overall background level in the image and we recommend that they be discarded. The default in \jude is to define the threshold as $m + 3\sigma$ where $m$ is the median of the counts per frame over the entire observation and $\sigma = \sqrt{m}$ for the photon counting UVIT detectors. If desired, the threshold may be changed using the max\_counts parameter (Table~\ref{tab:params}). Note that we record no more than 1000 events per frame, a most unlikely level for any observation to reach. 

Our final step in the production of Level 2 data was to remove duplicate frames and merge files which were broken in the middle of an exposure. This creates a number of FITS binary tables (Table~\ref{tab:datalevel2} which contain the time-tagged photon lists with selected housekeeping data. These can be used for further scientific processing without returning to the Level 1 data and the housekeeping files, saving considerable time. One of the most time-consuming aspects of \jude is matching the event lists with the housekeeping data.

\subsection{Level 2 Output Files}

\begin{table}[t]
\caption{Level 2 Data}
\label{tab:datalevel2}
\begin{tabular}{llll}
\hline
 & Variable & Data Type & Sample Value\\
\hline
1  &   FRAMENO     &  LONG   &   17007\\
2  &   ORIG\_INDEX &  LONG   &   17451\\
3  &   NEVENTS     &  INT    &   153\\
4  &   X           &  FLOAT  &   Array[1000]\\
5  &   Y           &  FLOAT  &   Array[1000]\\
6  &   MC          &  INT    &   Array[1000]\\
7  &   DM          &  INT    &   Array[1000]\\
8  &   TIME        &  DOUBLE &   1.8824165e+08\\
9  &   DQI         &  INT    &   0\\
10 &   FILTER		&	FLOAT & 271.232\\
11 &   ROLL\_RA    &  DOUBLE &  11.761461\\
12 &   ROLL\_DEC   &  DOUBLE &  85.226608\\
13 &   ROLL\_ROT   &  DOUBLE & 166.56405\\
14 &   ANG\_STEP   &  DOUBLE &   0.0030735851\\
15 &   XOFF        &  FLOAT  &  -0.998805\\
16 &   YOFF        &  FLOAT  &  -1.00348\\
\hline
\end{tabular}
\end{table}

Our primary Level 2 data product is a UV photon list for every exposure (a contiguous set of UV data). This list is time-tagged and stored in a FITS binary table (Table~\ref{tab:datalevel2}). Each row of the FITS binary table represents a single 0.035 second frame identified by the time in seconds from Jan. 1, 2010. The $x$ and $y$ position of every photon in a frame are stored in two arrays (``X'' and ``Y'') and the spacecraft motion is stored in ``XOFF'' and ``YOFF''. The output files from \jude are described in \ref{app:Level2_UV}, including the VIS output and the UV photon lists and images.

\subsection{Image Production}

In principle, it is easy to add the individual photon events from the Level 2 UV files to form an image of the sky, as we have done in Fig. \ref{fig:vis_correction} (using {\em jude\_add\_frames.pro}). However, the quality of the images will be determined largely by how well we are able to correct for the spacecraft motion. The VIS images were intended to track stars in the field and, as discussed in Section~\ref{sec:vis_proc}, we have matched the calculated VIS offsets with the UV frames to populate ``XOFF'' and ``YOFF'' in Table~\ref{tab:datalevel2}.

If the VIS data are not available, we obtained a first estimate of the motion from the on-board star sensor which reports the spacecraft boresight (RA, Dec, Roll Angle) every 16 seconds (464 frames). We used these to shift and add frames and, in the same manner as the VIS data, found the shifts between successive frames. Most of the UV sources observed to date are relatively faint, largely to protect the detectors from damage, and we had to integrate over 100 -- 200 frames (3.5 -- 7 seconds) to get enough counts for {\it find.pro} to work and interpolate to find the shifts for each frame.

\begin{figure}[t]
\includegraphics[width=3in]{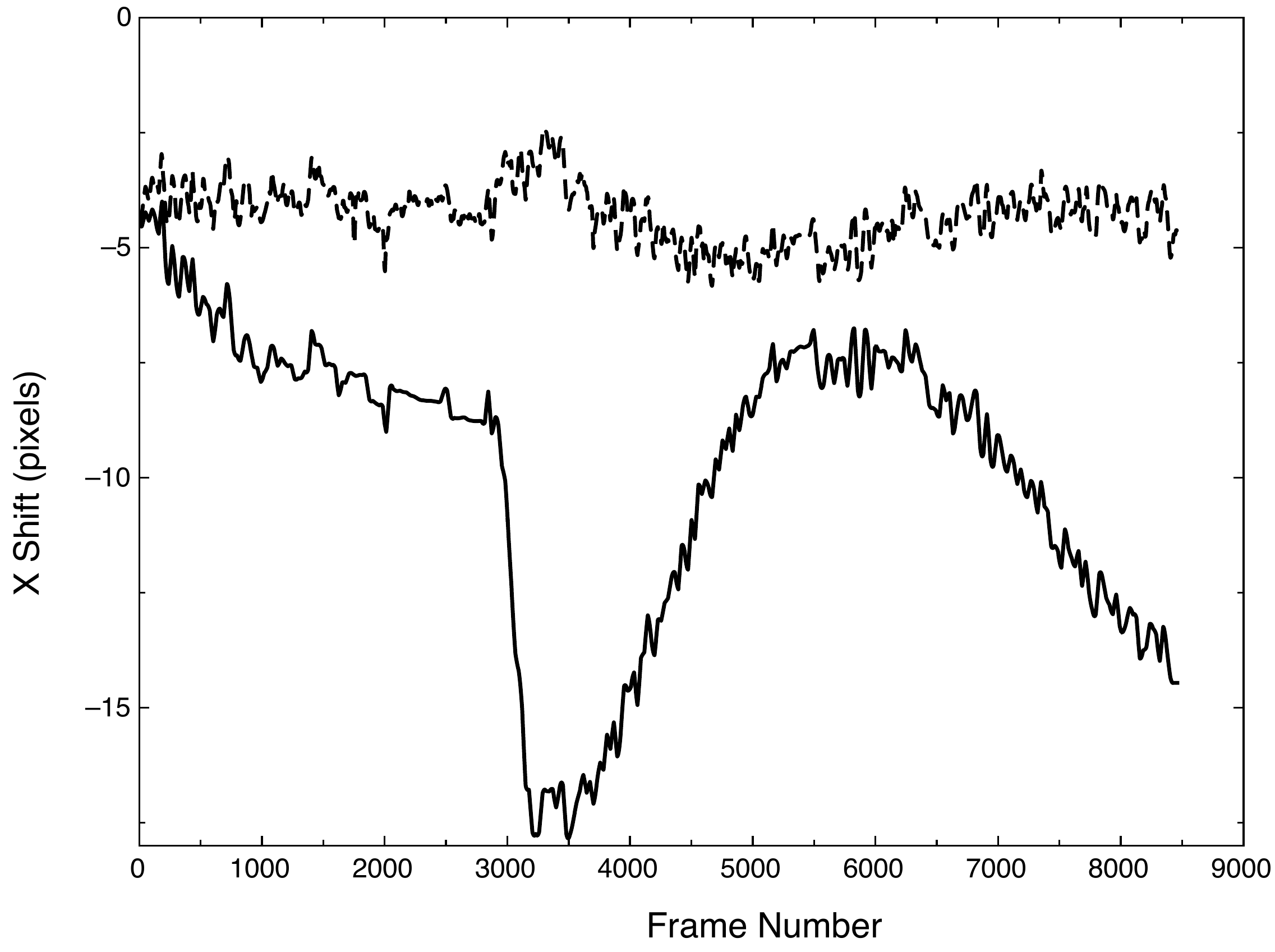}
\caption{Spacecraft motion derived from the VIS channel (solid line). The jerk at about frame 3000 is likely due to the operation of the Scanning Sky Monitor (SSM). The difference between the motion derived from the UV channel and the VIS channel is plotted as a dashed line. The offset of about 5 pixels is due to differing starting points.}
\label{fig:sc_motion}
\end{figure}

\begin{figure*}[t]
\includegraphics[width=3in]{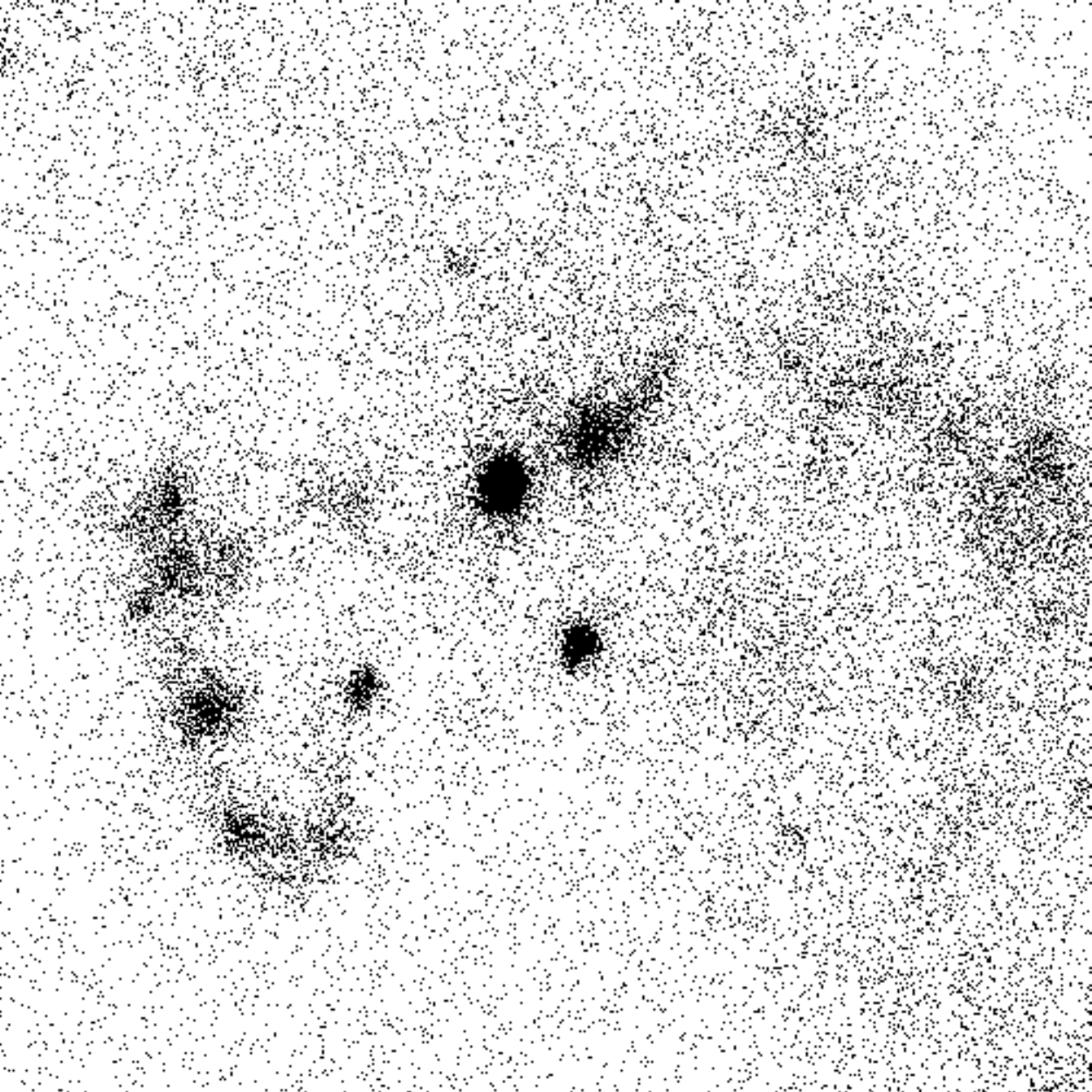}
\includegraphics[width=3in]{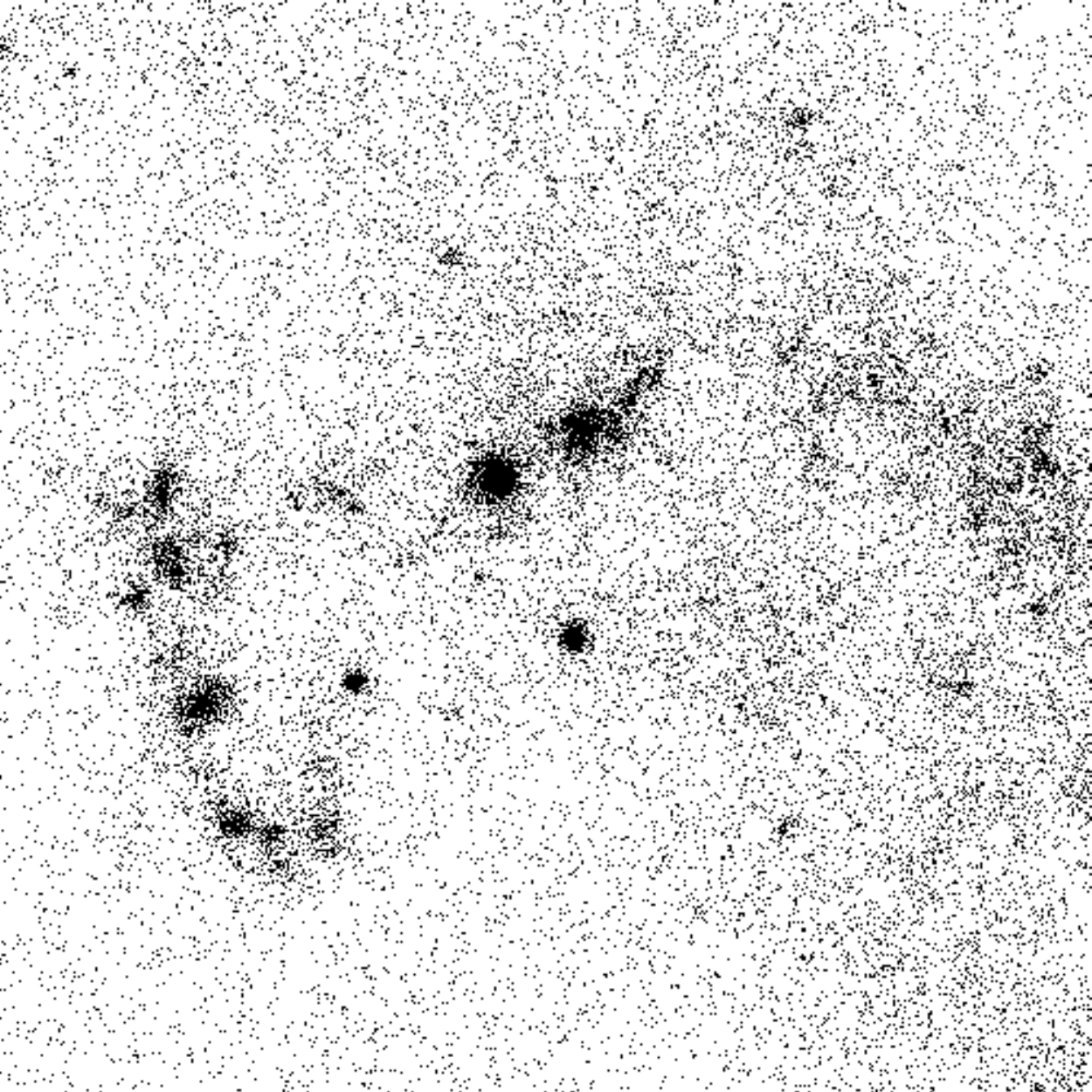}
\caption{We achieve better resolution when we use self-registration through {\it jude\_centroid} (right) than with the VIS-derived spacecraft motion (left).}
\label{fig:holm_res}
\end{figure*}

We found that we could dramatically improve the resolution if we adopted a more robust centroiding algorithm ({\it jude\_centroid.pro}) in which we followed a single star in the image throughout the observation. The registration worked best when we added together 10 -- 20 frames with a star that had 3 -- 10 counts in that time. We calculated the centroid of the star in $x$ and $y$ in each set of frames and used the shifts between frames to derive the spacecraft motion. The difference between the spacecraft motion derived here and that inferred from the VIS channel is plotted in Fig.~\ref{fig:sc_motion} and is on the order of 0.5 pixels (other than a constant offset of 5 pixels due to a different choice for the reference frame). The better resolution obtained with {\it jude\_centroid.pro} is apparent in Fig.~\ref{fig:holm_res} with more detail seen in the central portion of Holmberg~II. The effective FWHM of the stars is $1.7^{\prime\prime}$ when we use {\it jude\_centroid}, and $4.7^{\prime\prime}$ when we use the VIS offsets.

\section{Calibration}

We have used \jude to perform an in-flight calibration of UVIT \citep{Rahna2017} and have incorporated those results back in the programs. We have left the Level 2 data in units of counts per second but have added a field to the header with the scale factor for the appropriate conversion into energy units (ergs cm$^{2}$ s$^{-1}$ \AA$^{-1}$ (cps)$^{-1}$). We found no evidence for flat field variations, especially in view of the long star tracks due to the spacecraft motion. We have derived a distortion correction for the data but found that applying it did not improve the resolution. We therefore recommend that the distortion correction be done as part of the astrometric correction after the final image production.

The astrometric information in the header is based on the relatively poor attitude information from the star sensor. We made the decision that Astrometry.net \citep{Astrometry2010} has solved this problem and, in keeping with our philosophy of reusing existing components, recommend it to the user for precise astrometry.

\section{Conclusions}

We have designed a set of software routines that begin with Level~1 UVIT data from the ISSDC and produces images suitable for scientific purposes. We have tested the software on all our GT observations, and have produced Level~2 event lists and images for all observations through an automated process. A few observations required manual intervention, primarily due to registration problems. We have used {\em JUDE} to characterize the in-flight performance of UVIT \citep{Rahna2017} and now expect to move on to the scientific exploitation of the data.

We have released \jude under the Apache License 2.0,\footnote{https://www.apache.org/licenses/LICENSE-2.0.txt} and it is available on the Astrophysics Source Code Library \citep{Murthy_ascl} and on GitHub (\textit{https://github.com/jaymurthy/JUDE}). We are using data from the UVIT for our science and, therefore, will continue to maintain and improve {\em JUDE} as needed.

As a final note, the scientific exploitation of the UVIT data has been limited till now owing to the delays in the UVIT pipeline. We hope that with the release of {\em JUDE} to the astronomical community at large, more and better scientific results from the UVIT will be published. The only valid test of scientific software is if it is used widely, and we will continue to update the routines as we find errors or as they are reported to us. We welcome feature requests to improve the scientific utility of the programs.

\section*{Acknowledgements}

JM thanks John, Paul, George and Ringo for inspiration and the fortitude for dealing with the intricacies of dealing with a major mission. Many people at IIA, ISRO, IUCAA, TIFR, NRC (Canada) and Univ. of Calgary have contributed to different parts of the spacecraft, instrument and the operations. Drs. R. Mohan and J. Postma answered many questions about the data. We also acknowledge the Gnu Data Language (GDL), the IDL Astronomy Library and its many contributors. This research has made use of NASA Astrophysics Data System Bibliographic Services. We thank the referees for a thorough review that we believe has significantly improved the readability of the paper.

This research has been supported by the Department of Science and Technology under Grants No.~SR/S2/HEP-050/2012 dated 14-08-2013 to Christ University and EMR/2016/00145 to IIA.

\section*{References}
\bibliography{uvit_cal}

\appendix

\section{IDL External Modules} \label{app:external_modules}

\begin{table*}
\centering
\caption{External Library Routines}
\label{tab:idlastron}
\begin{tabular}{ll}
\hline
Module & Purpose\\
\hline
add\_distort.pro & Add the distortion parameters in an astrometry structure to a FITS header.\\
ad2xy.pro & Coordinate to pixel conversion.\\
cgErrorMsg.pro & a device-independent error messaging function. \\
correl\_images.pro & Compute the 2-D cross-correlation function of two images. \\
corrmat\_analyze.pro & Find the optimal ($x, y$) offset to maximize correlation of 2 images. \\
detabify.pro & Replace tabs with spaces.\\
data\_chk.pro & Checks input data.\\
datatype.pro & Returns data type of a variable.\\
daycnv.pro & Converts Julian dates to Gregorian calendar dates.\\
gcirc.pro & Computes rigorous great circle arc distances.\\
getcoords.pro & Converts a string with angular coordinates  to floating point values. \\
get\_date.pro & Return the UTC date in CCYY-MM-DD format.\\
get\_equinox.pro & Returns equinox.\\
getopt.pro & Convert a string into a valid scalar or vector.\\
gettok.pro & String manipulation.\\
get\_wrd.pro & Return the nth word from a text string.\\
file\_exist.pro & Tests for the existence of a file.\\
find.pro & Find point sources.\\
file\_stat.pro & Returns information on a file.\\
fxaddpar.pro & Add parameter to FITS header. \\
fxbhmake.pro & Create basic FITS header.\\
fxhclean.pro & Removes keywords from FITS header.\\
fxpar.pro & Obtains value of a parameter in the FITS header.\\
fxparpos.pro & Finds position to insert par. \\
fxposit.pro & FITS file manipulation.\\
fxmove.pro & Skip extensions in a FITS file.\\
make\_astr.pro & Builds astrometry structure.\\
match.pro & Find subscripts where vectors match.\\
mean.pro & Find mean.\\
mkhdr.pro & Makes FITS header.\\
mpfit.pro &  Perform Levenberg-Marquardt least-squares minimization.\\
mpfit2dfun.pro & Perform Levenberg-Marquardt least-squares fit to a 2-D IDL function. \\
mpfit2dpeak.pro & Fit a gaussian, lorentzian or Moffat model to data. \\
mrd\_hread.pro & Reads FITS header.\\
mrdfits.pro & Reads FITS files.\\
mrd\_skip.pro & Skip bytes in a file.\\
mrd\_struct.pro & Create structure.\\
mwrfits.pro & Writes FITS files.\\
putast.pro & Adds astrometry to FITS header.\\
quadterp.pro & Interpolation.\\
remove.pro & Remove elements from vector.\\
reverse.pro & Reverse the order of rows or columns in an array.\\
srcor.pro & Match point sources in image.\\
strsplit.pro & STRTOK wrapper.\\
sxaddhist.pro & Adds HISTORY to FITS header.\\
sxaddpar.pro & Adds parameter to FITS header.\\
sxpar.pro & Obtain the value of a parameter in a FITS header \\
tag\_exist.pro & Check for tags in a structure.\\
valid\_num.pro & Checks that a string is a valid number representation.\\
wcs\_check\_ctype.pro & Checks CTYPE parameters and return the projection type and coordinate type.\\
wcs\_rotate.pro & Rotate between standard and native coordinates.\\
wcssph2xy.pro & Coordinate to pixel conversion.\\
write\_png.pro & Writes PNG (Portable Network Graphics) file.\\
zparcheck.pro & Check type and size of a parameter.\\
\hline
\end{tabular}
\end{table*}

\jude uses routines from the IDL Astronomy Library \citep{Landsman1995} and from the MPFIT library \citep{mpfit2009}. These are listed in Table \ref{tab:idlastron} and may be downloaded individually or as part of the entire library.

\section{Running the pipeline} \label{app:running_the_pipeline}

We have written a GDL command file ({\it process\_uvit.com}) to call each of the programs in order. The user will have to change the first two lines to conform to their environment.
\begin{enumerate}
\item {\bf !path='jude\_dir:'+!path} --- {\it !path} is a system variable specifying the directories where the programs are located, if not in the working directory. It should include {\em JUDE} files and any additional programs required.
\item {\bf dname='root\_dir/'} --- Sets the variable {\it dname} to the root directory for the UVIT Level~1 data files.
\item {\bf jude\_driver\_vis,dname} --- Processes the VIS files and derives the spacecraft motion. The VIS processing should occur before the two UV channels because the image correction comes from the VIS channel.
\item {\bf jude\_driver\_uv,dname,/nuv} --- Processes the NUV files to produce event lists.
\item {\bf jude\_driver\_uv,dname,/fuv} --- Processes the FUV files to produce event lists. The two UV routines may be run in either order.
\item {\bf jude\_uv\_cleanup,/nuv} --- Merges the NUV event lists, matches with offsets from the VIS channel, and produces images.
\item {\bf jude\_uv\_cleanup,/fuv} --- Merges the FUV event lists, matches with offsets from the VIS channel, and produces images.
\end{enumerate}

\section{Parameters} \label{app:params}

\begin{table*}
\caption{Program parameters}
\label{tab:params}
\centering
\begin{tabular}{lp{3.5in}l}
\hline
Parameter & Description & Default Value\\
\hline
resolution & Number of sub-pixels in one physical pixel (1, 2, 4 or 8). & 8\\
min\_counts & Frames with less counts are rejected (Fig. \ref{fig:event_histogram}). & 0\\
max\_counts & Frames with more counts are rejected. If 0, we use the $median + 3 \times \sqrt{median}.$& 0\\
min\_frame & Starting frame. In practice, the first frame where the DQI value is 0. & 0\\
max\_frame & Ending frame. If 0, last frame where DQI = 0.& 0\\
coarse\_bin & Number of frames binned to find stars in {\it jude\_register.pro}.  & 200\\
fine\_bin & Number of frames binned to find centroids in {\it jude\_centroid.pro}. & 20\\
ps\_threshold\_fuv & Threshold for FUV point sources in counts s$^{-1}$ pixel$^{-1}$. Used in {\it jude\_register}. & $3 \times 10^{-4}$\\
ps\_threshold\_nuv & Threshold for NUV point sources in counts s$^{-1}$ pixel$^{-1}$. Used in {\it jude\_register}.& $1.5 \times 10^{-3}$\\
flat\_field   &  Place holder if we want to use a flat field. Not implemented as yet. & "No flat field"\\
events\_dir   & Output directory for Level 2 event lists. Separated by channel (VIS, NUV, FUV). & "events/"\\
image\_dir   & Output directory for FITS images.  & "images/" \\
mask\_dir   & Holds mask files. Each mask is a FITS file containing an $512 \times 512$ array where the region of interest is marked with 1 and the other elements are 0. Masks are not used in the pipeline but the functionality is in {\it jude\_interactive.pro} and the masks are written by {\it jude\_mask.pro}.  & "masks/" \\
def\_nuv\_dir & Default directory for NUV channel. & "nuv/"\\
def\_fuv\_dir & Default directory for FUV channel. & "fuv/"\\
def\_vis\_dir & Default directory for VIS channel. & "vis/"\\
png\_dir	  & Output directory for PNG diagnostic images. & "png/ \\
vis\_L2\_dir  & Output directory for VIS files. & "vis\_files/"\\
vis\_off\_dir & Output directory for files with spacecraft motion. & "vis\_off/"  \\
vis\_add\_dir    & Output directory for VIS images. & "vis\_add/"\\
temp\_dir  & Directory for temporary files. & "jude\_temp/"\\
\hline\\
\end{tabular}
\end{table*}

The parameters used in \jude are listed in Table~\ref{tab:params} and are passed to the pipeline through a structure defined in {\it jude\_params.pro}. The advantage of using a structure is that it allows for further parameters to be added if required, without changing the program invocation.

\section{Level 2 Data Format}\label{app:Level2_UV}
\subsection{VIS Files}
\begin{enumerate}
\item A GDL save set containing three variables:
\begin{enumerate}
\item GRID --- Floating point array of dimensions $512 \times 512 \times NFRAMES$. The VIS data consist of individual CMOS frames with $512 \times 512$ pixels with a total number of frames in the Level 1 data denoted by $NFRAMES$.
\item TIMES --- Double precision array with $NFRAMES$ elements containing the UVIT time (the number of seconds since Jan. 1, 2010).
\item DATA\_HDR --- String array with the header information from the Level 1 VIS file. The format is the standard GDL format for FITS headers with one line for each keyword in the original FITS header.
\end{enumerate}
\item A text file containing the offsets between each frame in the VIS data and the first frame. The first line has two elements: the starting and ending time for that file and each successive line has 4 elements, with the first column being the reference time and the second being the time for that particular frame. The last two columns are the $X$ and $Y$ shifts of that frame with respect to the reference frame. The times are specified to 6 decimal places with the first 4 being significant, and the $X$ and $Y$ offsets are specified to 2 decimal places.
\item A FITS file with the co-added VIS data, including the correction for spacecraft motion.
\end{enumerate}

\subsection{UV Data}

\begin{enumerate}
\item A FITS binary table for each of the two UV bands containing the photon event list as a function of frame number with relevant housekeeping and attitude information (Table \ref{tab:datalevel2}). An optional second extension contains the offsets from the VIS data. The components of the binary table are as follows:
\begin{enumerate}
\item FRAMENO is the frame number from the Level~1 data. A single frame may run across multiple lines if there are more than 336 events recorded, usually due to a cosmic ray hit. There may also be duplicated frames.
\item ORIG\_INDEX is the row number (beginning with 0 as in the GDL convention) in the original data file (recorded in the FITS header). ORIG\_INDEX is $\geq$ than FRAMENO due to the repetition of frames in the Level 1 data.
\item NEVENTS is the total number of photon events in the frame with a maximum of value of 999. 
\item $X$ is an array contains the column number of each event up to NEVENTS ranging from 0 to 511.875.
\item $Y$ is an array contains the column number of each event up to NEVENTS ranging from 0 to 511.875.
\item MC is the minimum corner value in the $5\times 5$ on-board centroiding. 
\item DM is the (maximum - minimum) $5\times 5$ corner value. MC and DM are reported as they may be used for diagnostics but are not used in {\it JUDE}.
\item TIME is the elapsed mission time in seconds from Jan. 1, 2010.
\item DQI is a diagnostic flag set by the pipeline. The values are defined in Table~\ref{tab:dqi}.
\item FILTER is the angle of the filter wheel where the conversion from angle to filter is in {\it jude\_set\_dqi.pro}.
\item ROLL\_RA is the right ascension (RA) of the spacecraft boresight as reported by the star sensor.
\item ROLL\_DEC is the declination (DEC) of the spacecraft boresight as reported by the star sensor.
\item ROLL\_ROT is the roll angle of the spacecraft boresight as reported by the star sensor.
\item ANG\_STEP is the shift in the spacecraft pointing between one frame and the next in arcseconds.
\item XOFF is the shift in sensor pixels (where the CMOS sensor is a $512 \times 512$ element array) between the frame and the reference frame in the $X$ direction, where the reference frame is defined by the FITS header of the image file.
\item YOFF is the same as XOFF but in the $Y$ direction.
\end{enumerate}

\item A FITS image file for each UV band with two extensions. The first extension is the co-added data from a single Level~1 file in units of counts s$^{-1}$ pixel$^{-1}$ with the second extension containing the exposure time per pixel. Although the file contains astrometric information, it is only as accurate as the spacecraft star sensor. The scale factor to convert from counts s$^{-1}$ pixel$^{-1}$ into ergs cm$^{-2}$ s$^{-1}$ \AA$^{-1}$ is in the FITS header.
\item A PNG file for a quick look containing the scaled image, a plot of the DQI values, a histogram of the number of counts, and a plot of the spacecraft motion.
\item A text file with a comma separated observation log.
\end{enumerate}

\end{document}